\documentclass{cjaa}
\usepackage{graphicx}
\input{epsf.sty}
\input{psfig.sty}

\def\beq{\begin{equation}}
\def\enq{\end{equation}}
\def\beqn{\begin{eqnarray}}
\def\enqn{\end{eqnarray}}

\begin{document}
\title{Millisecond pulsar population in the Galactic center and high energy contributions}

\author{Wei Wang\inst{1,2}}
\institute{National Astronomical Observatories, Chinese Academy of Sciences, Beijing 100012\\
\and Max-Planck-Institut f\"ur extraterrestrische Physik, Postfach
1312, 85741 Garching, Germany}

\date{Received~~2005; accepted~~2005}

\abstract{We propose that there possibly exists a population of
millisecond pulsars in the Galactic center region. Millisecond
pulsars (MSPs) could emit GeV gamma-rays through
synchrotron-curvature radiation as predicted by outer gap models.
In the same time, the compact wind nebulae around millisecond
pulsars can emit X-rays though synchrotron radiation and TeV
photons through inverse Compton processes. Besides, millisecond
pulsar winds provide good candidates for the electron-positrons
sources in the Galactic center. Therefore, we suggest that the
millisecond pulsar population could contribute to the weak
unidentified Chandra X-ray sources, the diffuse gamma-rays
detected by EGRET, electron-positron annihilation lines and TeV
photons detected by HESS toward the Galactic center. }

\authorrunning{W. Wang}
\titlerunning{Millisecond pulsar population in the Galactic center}

\maketitle

\section{Motivations}
Millisecond pulsars are old pulsars which could have been members
of binary systems and been recycled to millisecond periods, having
formed from low mass X-ray binaries in which the neutron stars
accreted sufficient matter from either white dwarf, evolved main
sequence star or giant donor companions. The current population of
these rapidly rotating neutron stars may either be single (having
evaporated its companion) or have remained in a binary system. In
observations, generally millisecond pulsars have a period $< 20$
ms, with the dipole magnetic field $< 10^{10}$ G. According to the
above criterion, we select 133 millisecond pulsars from the ATNF
Pulsar Catalogue
\footnote{http://www.atnf.csiro.au/research/pulsar/psrcat/}.
Figure 1 shows the distribution of these MSPs in our Galaxy, and
they distribute in two populations: the Galactic field (1/3) and
globular clusters (2/3). In the Galactic bulge region, there are
four globular clusters, including the famous Terzon 5 in which 27
new millisecond pulsars were discovered (Ransom et al. 2005).

\begin{figure}
\centering
\includegraphics[angle=0,width=7cm]{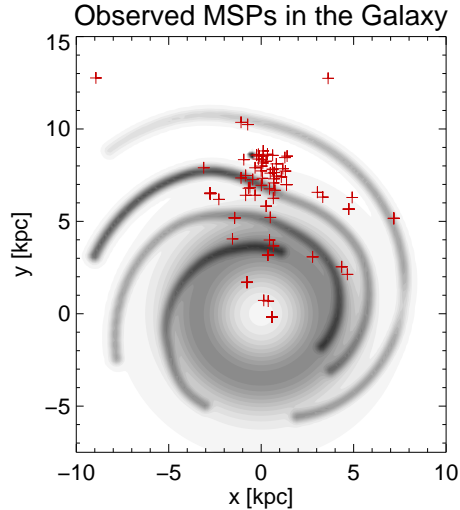}
\caption{The distribution of the observed millisecond pulsars in
the Milk Way. The grey contour is the electron density
distribution from Taylor \& Cordes (1993).}
\end{figure}

Recently, deep {\em Chandra} X-ray surveys of the Galactic center
(GC) revealed a multitude of point X-ray sources ranging in
luminosities from $\sim 10^{32} - 10^{35}$ ergs s$^{-1}$ (Wang,
Gotthelf, \& Lang 2002a) over a field covering a $ 2 \times 0.8$
square degree band and from $\sim 3 \times 10^{30} - 2 \times
10^{33}$ ergs s$^{-1}$ in a deeper, but smaller field of $17'
\times 17'$ (Muno et al. 2003). More than 2000 weak unidentified
X-ray sources were discovered in the Muno's field. The origin of
these weak unidentified sources is still in dispute. Some source
candidates have been proposed: cataclysmic variables, X-ray
binaries, young stars, supernova ejecta, pulsars or pulsar wind
nebulae.

EGRET on board the {\em Compton GRO} has identified a central
($<1^\circ$) $\sim 30 {\rm MeV}-10$ GeV continuum source (2EG
J1746-2852) with a luminosity of $\sim 10^{37}{\rm erg\ s^{-1}}$
(Mattox et al. 1996). Further analysis of the EGRET data obtained
the diffuse gamma ray spectrum in the Galactic center. The photon
spectrum can be well represented by a broken power law with a
break energy at $\sim 2$ GeV (see Figure 2, Mayer-Hasselwander et
al. 1998). Recently, Tsuchiya et al. (2004) have detected sub-TeV
gamma-ray emission from the GC using the CANGAROO-II Imaging
Atmospheric Cherenkov Telescope. Recent observations of the GC
with the air Cerenkov telescope HESS (Aharonian et al. 2004) have
shown a significant source centered on Sgr A$^*$ above energies of
165 GeV with a spectral index $\Gamma=2.21\pm 0.19$. Some models,
e.g. gamma-rays related to the massive black hole, inverse Compton
scattering, and mesonic decay resulting from cosmic rays, are
difficult to produce the hard gamma-ray spectrum with a sharp
turnover at a few GeV. However, the gamma-ray spectrum toward the
GC is similar with the gamma-ray spectrum emitted by middle-aged
pulsars (e.g. Vela and Geminga) and millisecond pulsars (Zhang \&
Cheng 2003; Wang et al. 2005a).

So we will argue that there possibly exists a pulsar population in
the Galactic center region. Firstly, normal pulsars are not likely
to be a major contributor according to the following arguments.
the birth rate of normal pulsars in the Milky Way is about 1/150
yr (Arzoumanian, Chernoff, \& Cordes 2002). As the mass in the
inner 20 pc of the Galactic center is $\sim 10^8 {\rm ~M}_{\odot}$
(Launhardt, Zylka, \& Mezger 2002), the birth rate of normal
pulsars in this region is only $10^{-3}$ of that in the entire
Milky Way, or $\sim$ 1/150 000 yr. We note that the rate may be
increased to as high as $\sim 1/15000$ yr in this region if the
star formation rate in the nuclear bulge was higher than in the
Galactic field over last $10^7 - 10^8$ yr (see Pfahl et al. 2002).
Few normal pulsars are likely to remain in the Galactic center
region since only a fraction ($\sim 40\%$) of normal pulsars in
the low velocity component of the pulsar birth velocity
distribution (Arzoumanian et al. 2002) would remain within the 20
pc region of the Galactic center studied by Muno et al. (2003) on
timescales of $\sim 10^5$ yrs. Mature pulsars can remain active as
gamma-ray pulsars up to 10$^6$ yr, and have the same gamma-ray
power with millisecond pulsars (Zhang et al. 2004; Cheng et al.
2004), but according to the birth rate of pulsars in the GC, the
number of gamma-ray mature pulsars is not higher than 10.

On the other hand, there may exist a population of old neutron
stars with low space velocities which have not escaped the
Galactic center (Belczynski \& Taam 2004). Such neutron stars
could have been members of binary systems and been recycled to
millisecond periods, having formed from low mass X-ray binaries in
which the neutron stars accreted sufficient matter from either
white dwarf, evolved main sequence star or giant donor companions.
The current population of these millisecond pulsars may either be
single or have remained in a binary system. The binary population
synthesis in the GC (Taam 2005, private communication) shows more
than 200 MSPs are produced through recycle scenario and stay in
the Muno's region.

\section{Contributions to high energy radiation in the Galactic
Center} Millisecond pulsars could remain active as high energy
sources throughout their lifetime after the birth. Thermal
emissions from the polar cap of millisecond pulsars contribute to
the soft X-rays ($kT < 1$ keV, Zhang \& Cheng 2003). Millisecond
pulsars could be gamma-ray emission source (GeV) through the
synchro-curvature mechanism predicted by outer gap models (Zhang
\& Cheng 2003). In the same time, millisecond pulsars can have
strong pulsar winds which interact with the surrounding medium and
the companion stars to produce X-rays through synchrotron
radiation and possible TeV photons through the inverse Compton
scatterings (Wang et al. 2005b). This scenario is also supported
by the Chandra observations of a millisecond pulsar PSR B1957+20
(Stappers et al. 2003). Finally, millisecond pulsars are potential
positron sources which are produced through the pair cascades near
the neutron star surface in the strong magnetic field (Wang et al.
2005c). Hence, if there exists a millisecond pulsar population in
the GC, these unresolved MSPs will contribute to the high energy
radiation observed toward the GC: unidentified weak X-ray sources;
diffuse gamma-ray from GeV to TeV energy; 511 keV emission line.
In this section, we will discuss these contributions separately.

\begin{figure}
\centering
\includegraphics[angle=0,width=10cm]{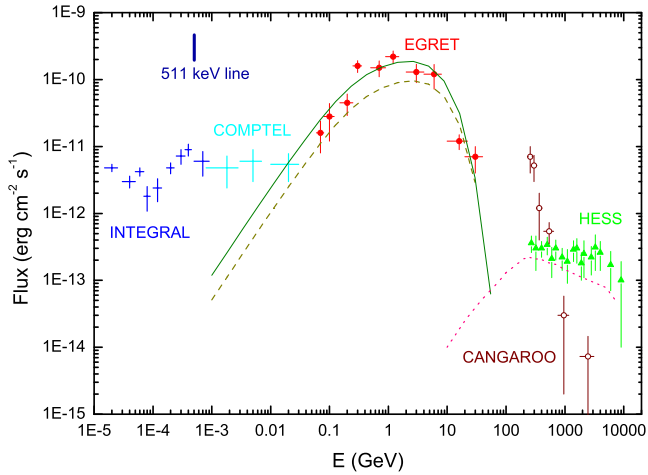}
\caption{The diffuse gamma-ray spectrum in the Galactic center
region within 1.5$^\circ$ and the 511 keV line emission within
6$^\circ$. The INTEGRAL and COMPTEL continuum spectra are from
Strong (2005), the 511 keV line data point from Churazov et al.
(2005), EGRET data points from Mayer-Hasselwander et al. (1998),
HESS data points from Aharonian et al. (2004), CANGAROO data
points from Tsuchiya et al. (2004). The solid and dashed lines are
the simulated spectra of 6000 MSPs according to the different
period and magnetic field distributions in globular clusters and
the Galactic field respectively. The dotted line corresponds to
the inverse Compton spectrum from MSPs.}
\end{figure}

\subsection{Weak unidentified Chandra X-ray sources}
More than 2000 new weak X-ray sources ($L_x>3\times 10^{30} {\rm
erg\ s^{-1}}$) have been discovered in the Muno's field (Muno et
al. 2003). Since the thermal component is soft ($kT < 1$ keV) and
absorbed by interstellar gas for sources at the Galactic center,
we only consider the non-thermal emissions from pulsar wind
nebulae are the main contributor to the X-ray sources observed by
Chandra (Cheng, Taam, Wang 2005). Typically, these millisecond
pulsar wind nebulae have the X-ray luminosity (2-10 keV) of
$10^{30-33} {\rm erg\ s^{-1}}$, with a power-law spectral photon
index from 1.5-2.5.

According to a binary population synthesis in the Muno's field,
about 200 MSPs are produced through the recycle scenario and stay
in the region if assuming the total galactic star formation rate
(SFR) of $1 M_\odot {\rm yr^{-1}}$ and the contribution of
galactic center region in star formation of 0.3\%. the galactic
SFR may be higher than the adopted value by a factor of a few
(e.g. Gilmore 2001), and the contribution of the galactic center
nuclear bulge region may be also be larger than the adopted values
(Pfahl et al. 2002). Then the actual number of MSPs in the region
could increase to 1000 (Taam 2005, private communication). So the
MSP nebulae could be a significant contributor to these
unidentified weak X-ray sources in the GC. In addition, we should
emphasize that some high speed millisecond pulsars ($>100$km\
s$^{-1}$) can contribute to the observed elongated X-ray features
(e.g. four identified X-ray tails have $L_x\sim 10^{32-33} {\rm
erg\ s^{-1}}$ with the photon index $\Gamma\sim 2.0$, see Wang et
al. 2002b; Lu et al. 2003; Sakano et al. 2003) which are the good
pulsar wind nebula candidates.

\subsection{Diffuse gamma-rays from GeV to TeV}
To study the contribution of millisecond pulsars to the diffuse
gamma-ray radiation from the Galactic center, e.g. fitting the
spectral properties and total luminosity, we firstly need to know
the period and surface magnetic field distribution functions of
the millisecond pulsars which are derived from the observed pulsar
data in globular clusters and the Galactic field (Wang et al.
2005a). We assume the number of MSPs, $N$, in the GC within $\sim
1.5^\circ$, each of them with an emission solid angle $\Delta
\Omega \sim$ 1 sr and the $\gamma$-ray beam pointing in the
direction of the Earth. Then we sample the period and magnetic
filed of these MSPs by the Monte Carlo method according to the
observed distributions of MSPs in globular clusters and the
Galactic field separately. We first calculate the fraction size of
outer gaps: $f\sim 5.5P^{26/21}B_{12}^{-4/7}$. If $f < 1$, the
outer gap can exist and then the MSP can emit high energy
$\gamma$-rays. So we can give a superposed spectrum of $N$ MSPs to
fit the EGRET data and find about 6000 MSPs could significantly
contribute to the observed GeV flux (Figure 2). The solid line
corresponds to the distributions derived from globular clusters,
and the dashed line from the Galactic field.

We can also calculate the inverse Compton scattering from the wind
nebulae of 6000 MSPs which could contribute to the TeV spectrum
toward the GC. In Figure 2, the dotted line is the inverse Compton
spectrum, where we have assumed the typical parameters of MSPs,
$P=3$ ms, $B=3\times 10^8$ G, and in nebulae, the electron energy
spectral index $p=2.2$, the average magnetic field $\sim 3\times
10^{-5}$ G. We predict the photon index around TeV:
$\Gamma=(2+p)/2=2.1$, which is consistent with the HESS spectrum,
but deviates from the CANGAROO data.

\subsection{511 keV emission line}
The Spectrometer on the International Gamma-Ray Astrophysical
Laboratory (SPI/INTEGRAL) detected a strong and extended
positron-electron annihilation line emission in the GC. The
spatial distribution of 511 keV line appears centered on the
Galactic center (bulge component), with no contribution from a
disk component (Teegarden et al. 2005; Kn\"odlseder et al. 2005;
Churazov et al. 2005). Churazov et al. (2005)'s analysis suggested
that the positron injection rate is up to $10^{43}\ e^+{\rm
s^{-1}}$ within $\sim 6^\circ$. The SPI observations present a
challenge to the present models of the origin of the galactic
positrons, e.g. supernovae. Recently, Cass\'e et al. (2004)
suggested that hypernovae (Type Ic supernovae/gamma-ray bursts) in
the Galactic center may be the possible positron sources.
Moreover, annihilations of light dark matter particles into
$e^\pm$ pairs (Boehm et al. 2004) have been also proposed to be
the potential origin of the 511 keV line in the GC.

It has been suggested that millisecond pulsar winds are positron
sources which result from $e^\pm$ pair cascades near the neutron
star surface in the strong magnetic field (Wang et al. 2005c). And
MSPs are active near the Hubble time, so they are continuous
positron injecting sources. For the typical parameters, $P=3$ ms,
$B=3\times 10^8$ G, the positron injection rate
$\dot{N}_{e^\pm}\sim 5\times 10^{37}{\rm s^{-1}}$ for a
millisecond pulsar (Wang et al. 2005c). Then how many MSPs in this
region? In \S 2.2, 6000 MSPs can contribute to gamma-rays with
1.5$^\circ$, and the diffuse 511 keV emission have a size $\sim
6^\circ$. We do not know the distribution of MSPs in the GC, so we
just scale the number of MSPs by $6000\times
(6^\circ/1.5^\circ)^2\sim 10^5$, where we assume the number
density of MSPs may be distributed as $\rho_{MSP}\propto
r_c^{-1}$, where $r_c$ is the scaling size of the GC. Then a total
positron injection rate from the millisecond pulsar population is
$\sim 5\times 10^{42}$ e$^+$ s$^{-1}$ which is consistent with the
present observational constraints. What's more, our scenario of a
millisecond pulsar population as possible positron sources in the
GC has some advantages to explain the diffuse morphology of 511
keV line emissions without the problem of the strong turbulent
diffusion which is required to diffuse all these positrons to a
few hundred pc, and predicts the line intensity distribution would
follow the mass distribution of the GC, which may be tested by
future high resolution observations.

\section{Summary}
In the present paper, we propose that there exists three possible
MSP populations: globular clusters; the Galactic field; the
Galactic Center. The population of MSPs in the GC is still an
assumption, but it seems reasonable. Importantly, the MSP
population in the GC could contribute to some high energy
phenomena observed by present different missions. A MSP population
can contribute to the weak unidentified Chandra sources in the GC
(e.g. more than 200 sources in the Muno's field), specially to the
elongated X-ray features. The unresolved MSP population can
significantly contribute to the diffuse gamma-rays detected by
EGRET in the GC, and possibly contribute to TeV photons detected
by HESS. Furthermore, MSPs in the GC or bulge could be the
potential positron sources. Identification of a millisecond pulsar
in the GC would be interesting and important. However, because the
electron density in the direction of the GC is very high, it is
difficult to detect millisecond pulsars by the present radio
telescopes. At present, we just suggest that X-ray studies of the
sources in the GC would probably be a feasible method to find
millisecond pulsars by {\em Chandra} and {\em XMM-Newton}.

\begin{acknowledgements}
W. Wang is grateful to K.S. Cheng, Y.H. Zhao, Y. Lu, K.
Kretschmer, R. Diehl, A.W. Strong, R. Taam, and the organizers of
this conferences at Hanas August 2005. This work is supported by
the National Natural Science Foundation of China under grant
10273011 and 10573021.
\end{acknowledgements}

\end{document}